\documentclass[12pt]{spieman}  
\usepackage{amsmath,amsfonts,amssymb}
\usepackage{graphicx}
\usepackage{setspace}
\usepackage{tocloft}
\usepackage{lineno}
\usepackage{soul}
\usepackage{amsmath}
\usepackage{xcolor}

\title{X-ray microcomputed tomography of 3D chaotic microcavities}

\author[a, b]{Ke Tian}
\author[a]{Mohammed Zia Jalaludeen}
\author[c]{Yeon Ui Lee}
\author[a, d, *]{Shilong Li}
\author[a,$\dagger$]{S\'ile Nic Chormaic}
\affil[a]{Okinawa Institute of Science and Technology Graduate University, Onna, Okinawa 904-0495, Japan}
\affil[b]{College of Physics and Optoelectronic Engineering, Harbin Engineering University, Harbin 150001, China}
\affil[c]{Department of Physics, Chungbuk National University, Cheongju, Chungbuk 28644, South Korea}
\affil[d]{College of Information Science and Electronic Engineering, Zhejiang University, Hangzhou 310058, China}

\cftpagenumbersoff{figure}
\cftpagenumbersoff{table} 
\begin{document} 
\maketitle

\begin{abstract}
Chaotic microcavities play a crucial role in several research areas, including the study of unidirectional microlasers, nonlinear optics, sensing, quantum chaos, and non-Hermitian physics. To date, most theoretical and experimental explorations have focused on two-dimensional (2D) chaotic dielectric microcavities, while there have been minimal studies on three-dimensional (3D) ones since precise geometrical information of a 3D microcavity can be difficult to obtain. Here, we image 3D microcavities with submicron resolution using X-ray microcomputed tomography ($\mu$CT), enabling nondestructive imaging that preserves the sample for subsequent use. By analyzing the ray dynamics of a typical deformed microsphere, we demonstrate that a sufficient deformation along all three dimensions can lead to chaotic ray trajectories over extended time scales. Notably, using the X-ray $\mu$CT reconstruction results, the phase space chaotic ray dynamics of a deformed microsphere are accurately established. X-ray $\mu$CT could become a unique platform for the characterization of such deformed 3D microcavities by providing a precise means for determining the degree of deformation necessary for potential applications in ray chaos and quantum chaos. 
\end{abstract}


{\noindent \footnotesize\textbf{*}Shilong Li,  \linkable{shilong.li@oist.jp} }
{\noindent \footnotesize\textbf{$\dagger$}S\'ile Nic Chormaic,  \linkable{sile.nicchormaic@oist.jp} }

\begin{spacing}{2}   

\section{Introduction}
Optical microcavities are capable of storing light energy in a small spatial region, thereby significantly enhancing light-matter interactions\cite{xiao2020ultra,spontaneous}. For optical frequencies, microcavities made from transparent dielectric materials have very low loss, and whispering gallery resonators (WGRs) have attracted significant research interest in linear and nonlinear optics, and quantum optics. Due to their ultra-high quality ($Q$) factor and small mode volume, WGRs have already served as excellent platforms in diverse applications such as sensing \cite{RingupNC,LanYangNature, YYang2016, YunfengLight, DeshuiYu, Frigenti23}, optical trapping and manipulation \cite{Jonathanoptica, LHogan2019, CouplingPRL,FPan2022}, microlasers \cite{YWu2010,YunfengLPR, Fang:17}, cavity quantum electrodynamics (QED) \cite{QEDOL,QEDPRA}, optical frequency combs \cite{Yang:16,Science, Armanicomb, Bluecomb,FuchuanOptica}, plasma and microwave photonics \cite{microwave, Bathish2023}. Due to the relative ease in cavity fabrication and characterization, (quasi) rotationally symmetric WGRs, which can achieve high $Q$-factors, are widely used in the above applications. However, asymmetric (that is deformed) WGRs can exhibit unique and complex physical phenomena arising from the chaotic ray trajectories that can arise under certain conditions, thereby driving research into chaotic dielectric microcavities in recent years.

Deformed WGRs have become a key system for the study of nonlinear dynamics, photon transport, quantum chaos, and non-Hermitian physics \cite{YunfengxiaoPRL,eLight,SileAPL}. For example, deformed microdisk cavities were used for unidirectional laser emission in the far-field \cite{CaoPRL}, effective microlaser collection \cite{QinghaiPRL}, and chaos-assisted tunneling (CAT) \cite{QinghaiLight}. Broadband and fast momentum transformation, as well as broadband optical frequency combs, were demonstrated in deformed microtoroid cavities \cite{YunfengScience,microcombYunfeng_NC}. Such works have mainly focused on two-dimensional (2D) deformed microcavities since the degree of deformation can be relatively well-defined and characterized by optical microscopy or scanning electron microscopy (SEM) \cite{microdiskfabrication}. In contrast, three-dimensional (3D) deformed microcavities have more degrees of freedom; such 3D chaotic cavities are ideal model systems for the study of wave dynamics \cite{natureray}, nonlinear optics \cite{YunfengNP}, strongly coupled cavity QED \cite{HailinWang,NV_microsphere}, optomechanics \cite{2016Yangchaosoptomechanics,2024PRAchaosoptomechanics}, and non-Hermitian physics \cite{caoreview}. These degrees of freedom are essential for exploring more realistic and complex ray dynamics, and are a better representation of naturally occurring optical systems. Furthermore, 3D microcavities are indispensable for studying full 3D wave chaos, optomechanical coupling, and isotropic quantum field interactions, all of which cannot be fully captured by simplified 2D models. However, the degree of deformation of a 3D microcavity is difficult to accurately control during the preparation stage, thereby requiring subsequent geometric characterization \cite{PRshilong}. Unfortunately, effective and nondestructive structural characterization methods for 3D chaotic microcavities, used to establish complex ray dynamics in phase space, are currently absent and this hinders their applications in fundamental physics. 

Here, we demonstrate a method for non-destructive imaging of 3D deformed dielectric microcavities by X-ray microcomputed tomography ($\mu$CT). Leveraging the precise geometric information obtained from the X-ray $\mu$CT reconstruction, we accurately establish the chaotic ray dynamics in phase space. In Section 2.1, we describe the experimental method to fabricate a deformed, asymmetric silica microsphere as an example of a chaotic dielectric microcavity. The geometric quantification of the cavity's profile, hence its deformation in all three spatial dimensions, via X-ray $\mu$CT is explained in detail in Section 2.2. Furthermore, discussions on the imaging resolution and voxel size of the X-ray-based 3D microcavity measurement system are presented, showing submicron resolution, capable of the geometrical characterization of most existing 3D chaotic microcavities. Finally, in Section 2.3, we present a discussion on chaotic ray dynamics that can appear in deformed microspheres (3D) through real- and phase-space diagrams and show that clear control over the deformation is necessary in order to fully reveal the chaotic trajectories. Our work provides an inspiring route to nondestructively image 3D microcavities and should advance the development of fundamental and applied physics in several areas, such as nonlinear dynamics, quantum chaos, and broad-area lasers.

\section{Experiment and Results}

\subsection{Fabrication of a 3D chaotic microcavity}
\label{sect:title}

\begin{figure*} [ht]
\centering
      \includegraphics[width=1.0\textwidth]{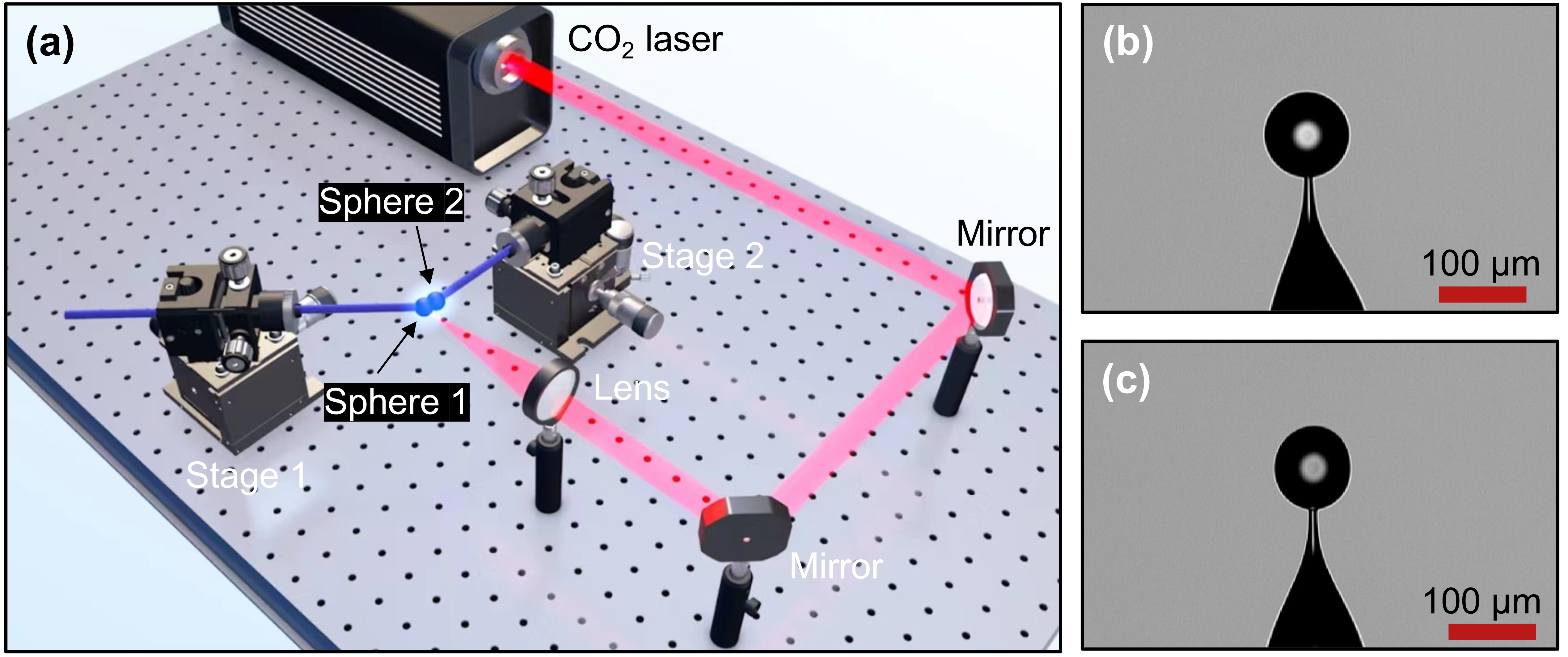}
  \caption{\textbf{Fabrication of deformed microspheres.} (a) Schematic diagram of the experimental setup. Optical microscope images of (b) a rotationally symmetric microsphere and (c) a deformed (asymmetric) microsphere.}
  \label{fig.1}
\end{figure*} 

A deformed silica microsphere was used as an example of a 3D chaotic dielectric microcavity. The experimental setup for the fabrication of the deformed microsphere  is illustrated in Fig.~\ref{fig.1}(a), where a focused CO$_{2}$ laser beam (48-2KWM, Synrad) was used to melt single mode fiber (Thorlabs, SM1550BHP); the focal length of the lens used was 75 mm. The CO$_{2}$ laser employed in this study operated at a wavelength of 10.6 $\mu$m in continuous-wave mode, with an output power set to 15\% of the maximum power of 10 W. The fabrication of rotationally symmetric microspheres from optical fiber is mature and has been widely reported \cite{Jonathanreview}. Typically, a rotationally symmetric (undeformed) microsphere can be obtained by melting the tip of a tapered optical fiber and the microsphere is retained on the optical fiber stem (see Fig. \ref{fig.1}(b)), facilitating subsequent manipulation. To make a deformed microsphere, we used a process very similar to that initially presented by Lacey and Wang \cite{microspherefabrication_OL}. First, two undeformed microspheres of similar sizes were fabricated and mounted on two 3D translation stages with their fiber stems perpendicular to each other. Next, the two microspheres were brought into contact with each other. Ideally, a deformed microsphere should only have one fiber stem as a support to ensure efficient excitation of whispering gallery modes (WGMs) and ray trajectories. To remove one of the two stems, we preheated the contact area of the two microspheres using the CO$_{2}$ laser to form a single fused microsphere structure. This enabled us to remove one fiber stem while retaining the other. Finally, the fused structure was slowly reheated until surface tension produced the desired degree of deformation. The deformation of the microspheres was primarily controlled by adjusting the laser heating duration, which was precisely programmed to ensure reproducible regulation of the deformation degree. During fabrication, a CCD camera  was used to capture real-time 2D projection images of the microspheres, allowing for direct observation of their geometric evolution. It is worth noting that if continuous heating were performed, the degree of deformation of the fused structure could approach zero. The aforementioned method is suitable for the fabrication of microspheres with large deformation, but for low deformation microspheres or on-chip microcavities, short CO$_{2}$ laser pulses are often used in post-processing \cite{YunfengScience, AM, AO}.

Figures \ref{fig.1}(b) and (c) show optical microscope images of a fabricated symmetric microsphere and deformed microsphere, respectively. The symmetric microsphere had a diameter of 92.6~$\mu$m and its rotational symmetry allowed it to be easily characterized using an optical microscope in a single plane. Even for asymmetric 2D microcavities, it is feasible to optically image the microcavity and fit a curve to its contour. However,  deformed microspheres do not have axial symmetry, making it difficult to determine their exact shape and degree of deformation. The size of deformed microspheres is usually defined by the long-axis dimensions. However, due to the lack of methods for effectively characterizing the entire geometry of 3D chaotic microcavities, their ray dynamics evolution can only rely on approximate treatments, such as dipole deformation \cite{microspherefabrication_OL}. This obstacle to quantify precisely the actual geometric parameters of 3D chaotic microcavities led to us explore an alternative technique, namely X-ray $\mu$CT.

\subsection{X-ray \texorpdfstring{$\mu$}{mu}CT characterization}
The X-ray $\mu$CT technique is well-known in medical and clinical applications \cite{Xrayreview}; however, it is also finding increasing scientific applications in biology and materials science in recent years \cite{xrayfiber,Parasitology2018,x-rayfiberOME}. X-ray $\mu$CT is a nondestructive technique for visualizing the external and internal structure of objects in 3D. Following the initial collection of a sample, no cutting or coating procedures are required \cite{non-destructive}. The scanning system produces 3D representations of a slice of an object based on material density, measured by X-ray transmissions. The slice is made up of voxels, which are the 3D equivalent of pixels. Each voxel is assigned a grey value derived from a linear attenuation coefficient that relates to the density of the material being scanned. Hence, a $\mu$CT scan is a mathematical representation of an object rather than a true image, making it well suited to access geometrical information about 3D chaotic microcavities. 

\begin{figure*}[ht]
\centering
      \includegraphics[width=1.0\textwidth]{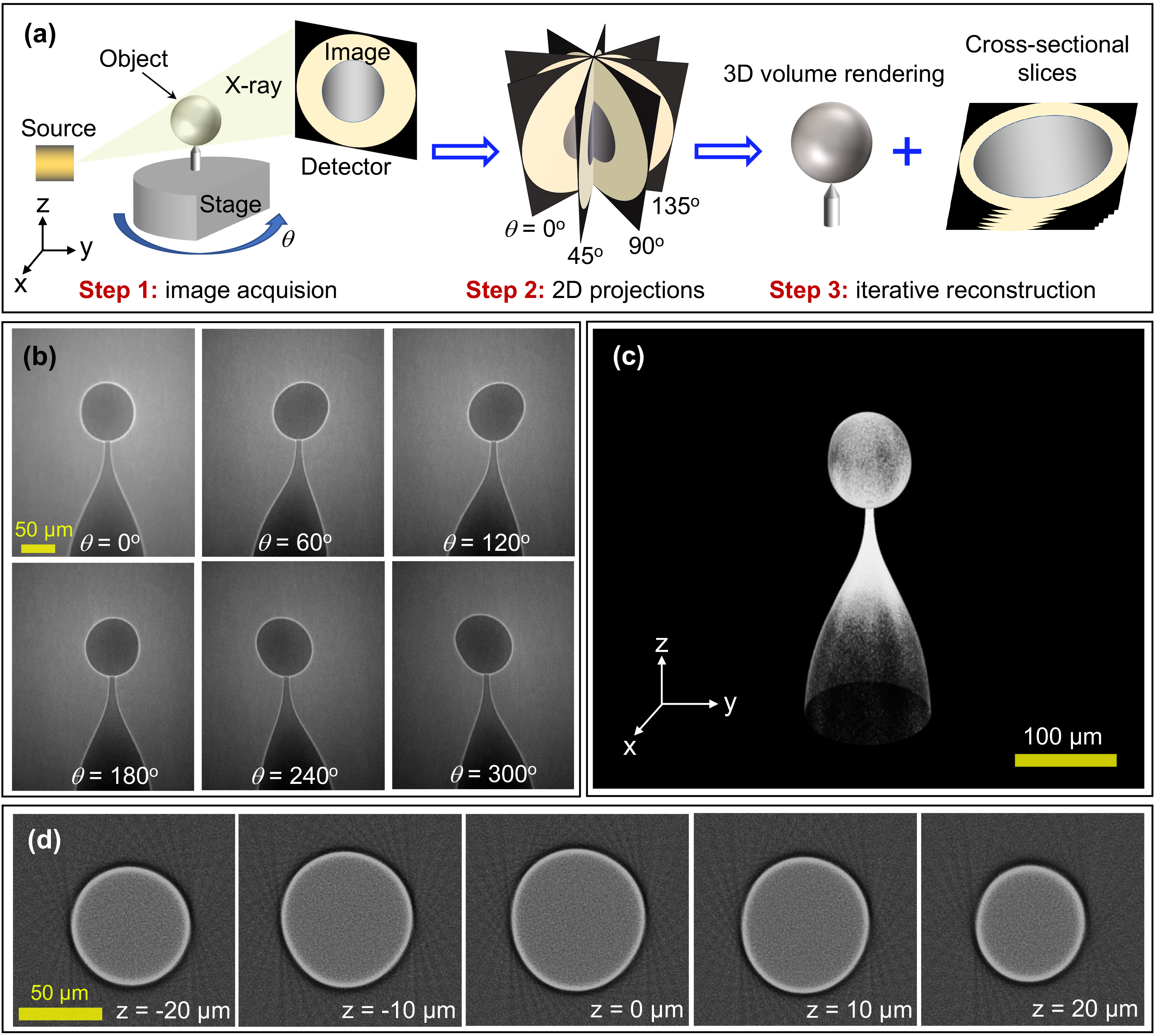}
\caption{\label{fig.2}\textbf{X-ray $\mu$CT characterization of a deformed microsphere.} (a) Illustration of the X-ray $\mu$CT setup and workflow. The projection images are obtained by rotating the sample in increments of $\theta$. The raw output of the tomogram is a series of projections of the sample taken at different viewing angles. The iterative reconstruction output can be visualized in both 3D rendered volume and 2D cross-sectional slices. (b) Projection images of the deformed microsphere at different $\theta$. (c) 3D rendered volume result of the deformed microsphere. (d) 2D cross-sectional slice results of the deformed microsphere at different positions along the $z$ axis.}
\end{figure*}

Figure \ref{fig.2}(a) illustrates the X-ray $\mu$CT setup (Zeiss Xradia 510 Versa) used in the experiment. It consists of three main elements: an X-ray source, a rotatable sample holder, and an X-ray detector. The X-rays were generated by the source and emitted toward the target sample. When passing through the sample, the X-rays were attenuated based on the properties of the sample being scanned (e.g., its density, thickness, and constituents). X-ray $\mu$CT characterization of a deformed microsphere consisted of three steps: image acquisition, 2D projections, and iterative reconstruction, as shown in Fig. \ref{fig.2}(a). The acquisition was completed by collecting 2D projection images (radiographs) from many viewing angles. Following their acquisition, the 3D volume images and cross-sectional images of the specimen were obtained by reconstructing these 2D projection images. 

In our measurements, the X-rays were generated from a molybdenum target using a voltage and current of 80~kV and 100--120~$\mu$A, respectively. A total of 721 angular projections were collected at 0.5\textdegree angular intervals in a single 360\textdegree  rotation. In Fig. \ref{fig.2}(b), six sample projection images of the deformed microsphere are depicted at 60\textdegree angular intervals. It is clear that the projections at varying angles exhibit differences, thereby rendering a rudimentary 2D characterization insufficient for a 3D chaotic microcavity. The radial projections were then reconstructed into a 3D matrix of isotropic voxels. Figure \ref{fig.2}(c) shows the 3D rendered volume result for the deformed microsphere. A 360\textdegree rotation animation of the deformed microsphere can be found in Visualization 1 of the supplementary material. 

For a solid deformed microsphere, the cross-sectional slice information at different positions along the $z$ axis may be more interesting, as it is crucial for selecting chaotic channels and constructing chaotic ray dynamics. Figure \ref{fig.2}(d) shows five cross-sectional reconstructed images selected at different positions on the $z$ axis ($z = 0$ is defined as being at the mid-point of the actual maximum size of the microsphere in the $z$ direction). As expected, the area and shape of the cross-sections of the deformed microsphere change along the $z$ axis. Based on the above tests, it is evident that X-ray $\mu$CT can effectively yield the geometrical information of 3D chaotic microcavities, including the volume geometry and the slice profile.

It has been proposed that the 3D structural information of a deformed microsphere could also be obtained by means of optical photogrammetry \cite{Li2022}. The proposed photogrammetry using visible light can only explore the external geometry of the microcavity, while X-ray tomography can access both the internal and external geometrical information of the microcavity, making it advantageous in the characterization of deformed microcavities with hollow structures, such as microbubble cavities and rolled-up cavities \cite{PRshilong,ShilongPRL,microbubbleliquid}. Moreover, there are challenges in obtaining high-quality 3D images from the proposed microcavity photogrammetry using visible light based on an optical microscope. Recall the Abbe diffraction limit for an optical microscope, given by $\Delta=\lambda/2\text{NA}$ where $\Delta$ is the spatial resolution of the microscope, $\lambda$ is the wavelength of light used, and NA is the numerical aperture of the microscope's objective. To observe (for obtaining 2D projection images) deformed microcavities around 100 $\mu$m, such as the asymmetric microsphere shown in Fig. \ref{fig.1}(c), typical objectives can be chosen with the following parameters: magnification of usually 4$\times$ or 10$\times$, NA of approximately between 0.1 and 0.25. Therefore, in the best scenario, microscopes suitable for microcavity photogrammetry have an upper limit for the resolution of 2D projection images of around 1~$\mu$m. In reality, however, unavoidable caustics will appear in these 2D projection images due to optical interference at the edges of the cavities; this would be especially severe for deformed microcavities. As a result, a realistic spatial resolution for the proposed microcavity photogrammetry is a few $\mu$m. Such a spatial resolution, which corresponds to an uncertainty of a few percent in the geometry, is far too large to study the chaotic dynamics of deformed microcavities. 

\begin{figure*}[ht]
\centering
      \includegraphics[width=1.0\textwidth]{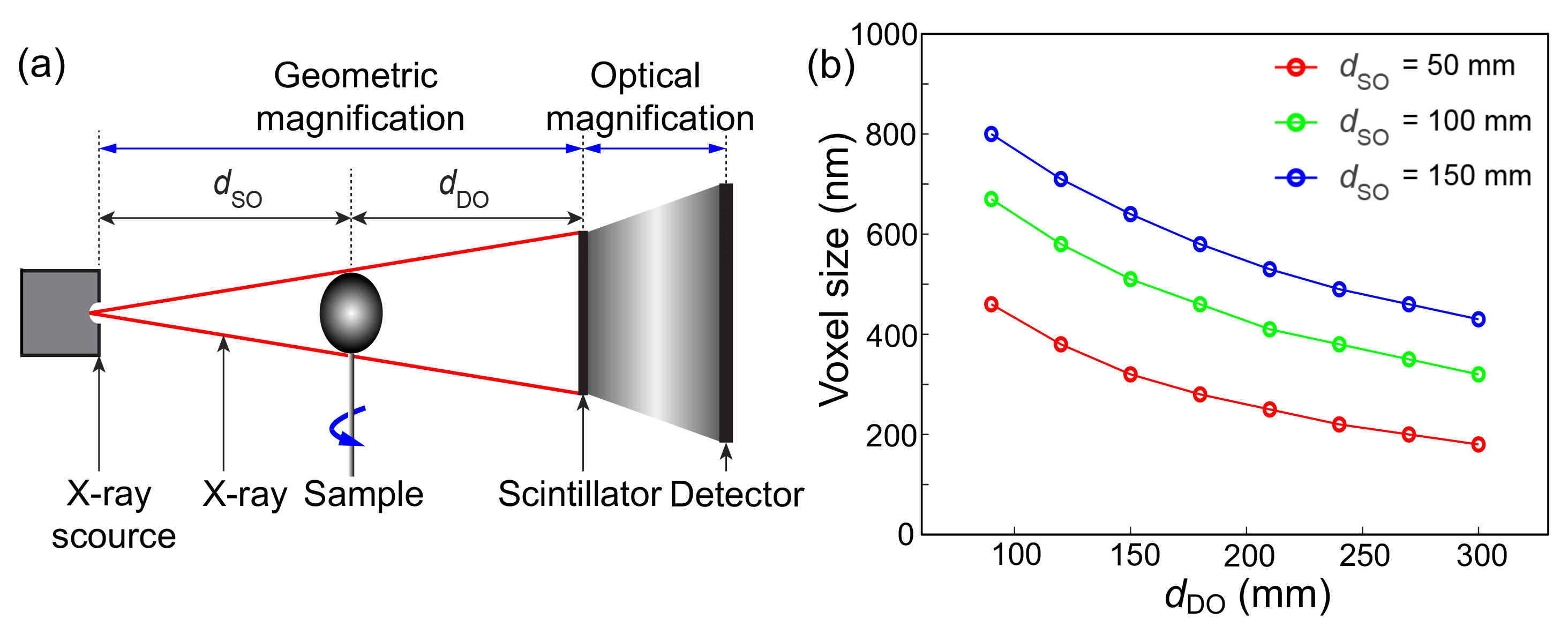}
\caption{\label{fig.3}\textbf{Dependence of the voxel size}. (a) Schematic diagram of the X-ray $\mu$CT system, showing its two-stage magnification technique. The object illuminated by the X-rays is a microsphere in our case. (b) Voxel size at various source-to-object distances ($d_{\text{SO}}$) and detector-to-object distances ($d_{\text{DO}}$). A 20$\times$ objective was used.}
\end{figure*}

Despite the lack of an X-ray microscope, the X-ray $\mu$CT system used in this work can achieve true sub-$\mu$m spatial resolution based on a two-stage magnification technique, as shown in Fig. \ref{fig.3}(a). This corresponds to a deformation uncertainty of less than 1 percent, enabling a feasible study of chaotic dynamics. There are two points to note. First, the voxel (also referred to as “nominal resolution”) is related to, but does not determine, the spatial resolution; nonetheless, the smaller the voxel, the better the spatial resolution achieved. Second, all of these resolutions, such as the spatial resolution and the voxel size, are functions of the spot size of the X-ray beam, the degree of the beam collimation, the size of the sample to be imaged, the size of the X-ray detector, the source-to-object distance ($d_{\text{SO}}$), the detector-to-object distance ($d_{\text{DO}}$), and the reconstruction algorithms. The voxel size depends on the detector pixel pitch and the geometric magnification of the system, which can be calculated by
\begin{equation}
\begin{split}
\label{Eq1}
    V=\frac{P}{M}, M=\frac{d_{\mathrm{SO}}+d_\mathrm{{DO}}}{d_{\mathrm{SO}}}  
\end{split}
\end{equation}
 where $V$ is the voxel size, $P$ is the detector pixel pitch, and $M$ is the geometric magnification of the system. Thus, moving the sample closer to the X-ray source or further away from the detector reduces the effective voxel size. This was experimentally verified, and the results are shown in Fig. \ref{fig.3}(b).

\subsection{Observation of 3D chaotic ray dynamics}

Once the 3D reconstruction results of an object’s X-ray $\mu$CT are obtained, the surface contour data can be extracted, and then a mathematical description of the surface can be derived through fitting methods (see surface fitting results in Appendix Fig. \ref{fig.App}). This allows us to determine the degree of deformation of a microcavity and the plane(s) of deformation. Specifically, the surface of the deformed microsphere produced by the approach described in Fig. \ref{fig.1} can be fitted using dipole and/or quadrupole functions. Using the methods in \cite{microspherefabrication_OL} we define two new parameters $u$ and $v$ such that: 
\begin{equation}
\begin{split}
\label{Eq2}
    \begin{cases}
    x(u,v)=&[1+e\text{cos}(2v)][1-q\text{cos}(2u)][1+d\text{cos}(u)\text{sin}(v)]\text{sin}(v)\text{cos}(u)/(1+q),\\
    y(u,v)=&[1+e\text{cos}(2v)][1-q\text{cos}(2u)][1+d\text{cos}(u)\text{sin}(v)]\text{sin}(v)\text{sin}(u)/(1+q),\\
    z(u,v)=&[1+e\text{cos}(2v)]\text{cos}(v).
    \end{cases}
\end{split}
\end{equation}

Here, the two parameters $0 \leq u < 2\pi$ and $0 \leq v < \pi$ (which differ from the spherical coordinates $\varphi$ and $\theta$, requiring a coordinate transformation) are used, where $u$ denotes the azimuthal angle (rotation around the z-axis in the x–y plane) and $v$ the polar angle (inclination from the north pole to the south pole). Together with the three variables $\{q, e, d\}$, they quantify the degree of deformation from an ideal spherical shape: ($\romannumeral1$) $e$ governs the quadrupolar deformation in the $y$-$z$ plane, significantly affecting the ray dynamics close to this plane; ($\romannumeral2$) $q$ determines the quadrupolar deformation in the $x$-$y$ plane, which is much smaller than $e$ and helps stabilize rays near the $y$-$z$ plane; ($\romannumeral3$) $d$ controls the dipole deformation in the $x$-$y$ plane, breaking the rotational symmetry around the $z$ axis and enabling chaotic 3D ray dynamics. Refractive escape of light from the cavity occurs when the angle of incidence, $\chi$, is equal to or larger than the critical angle, $\chi_c = \sin^{-1}(1/n)$ where $n$ is the refractive index of the resonator's medium. For fused silica, $\sin\chi_c = 0.69.$ The ray trajectories for the 3D deformed system can then be described in phase-space using a Poincar\'e surface of section (SOS), which shows the intersection of the trajectories across chosen planes in phase-space.

\begin{figure*}[ht]
\centering
      \includegraphics[width=0.85\textwidth]{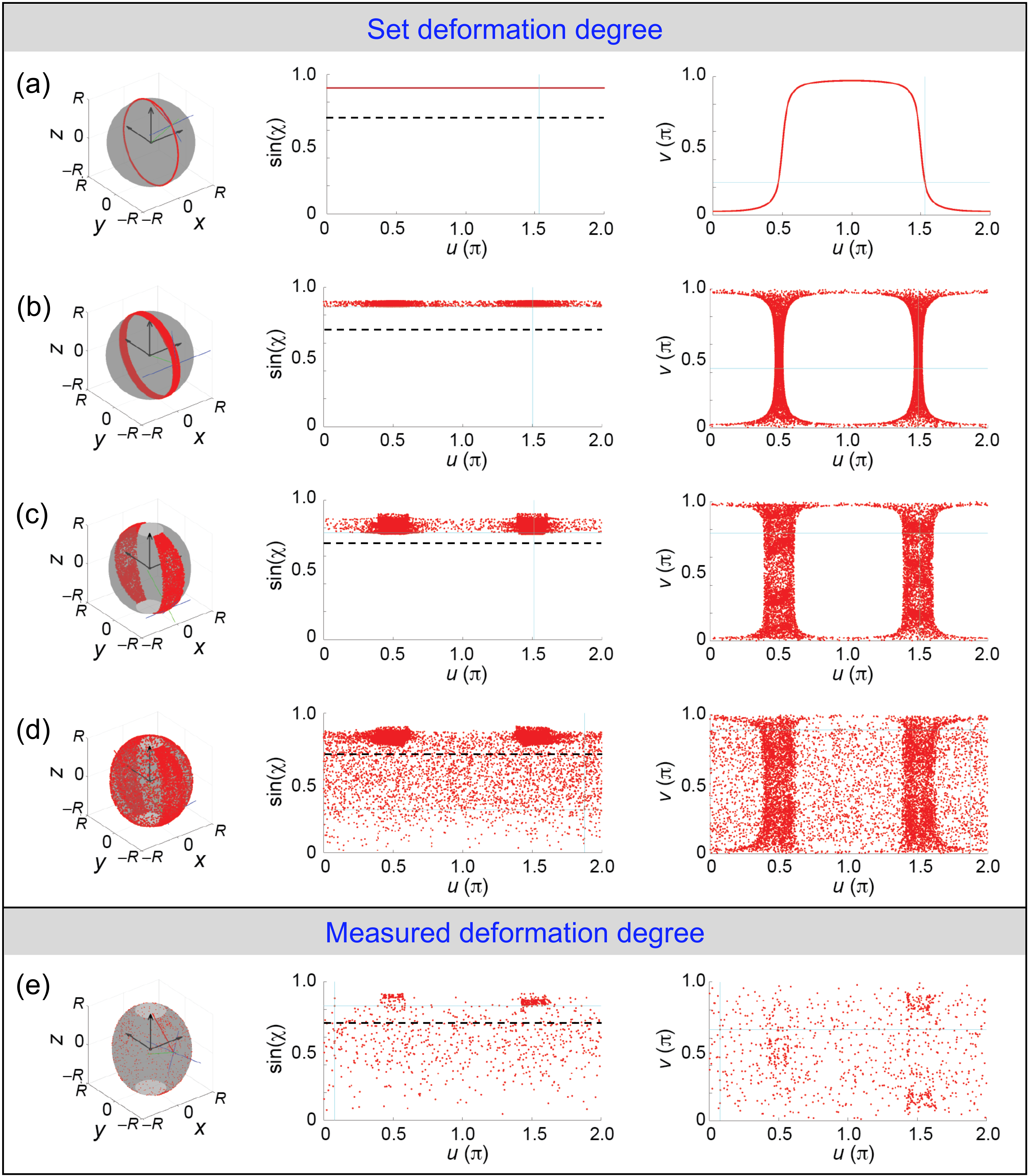}
\caption{\label{fig.4}\textbf{Computational results illustrating the evolution of ray trajectories in a deformed microsphere under varying deformation levels.}. (a--e) Ray trajectory in real-space (left panel) and two selective SOSs (middle and right panels). A single trajectory starting in the plane inclined ($\sim5$ degrees) from the vertical ($z$-axis) with $\text{sin}(\chi_0)=0.9+0.01*(\sqrt{2}-1)$ was followed for about 1,000,000 reflections in (a--d) and 100,000 reflections in (e), respectively. The deformation $\{q,e,d\}$ is $\{0,0,0\}$ in (a), $\{0.005,0,0\}$ in (b), $\{0.005,0.06,0\}$ in (c), $\{0.005,0.06,0.05\}$ in (d), and $\{0.02,0.08,0.05\}$ in (e), respectively. Black dashed line: critical line $\text{sin}(\chi_c)=1/n$.}
\end{figure*}

To illustrate these effects, a series of 3D ray tracing calculations for a deformed microsphere at various degrees of deformation were performed and the results are summarized in Fig. \ref{fig.4}(a--d). We plot both the ray trajectory in real-space and in two selected Poincar\'e SOSs; however, we only represent 2D plots for ease of viewing. The empty spaces in the SOS plots represent regions of phase-space that are forbidden due to the angular momentum barrier, $L_z$, along the $z$-direction \cite{1995PRL}. Note that Fig. \ref{fig.4}(a--d) is not based on the actual fitting results to our deformed microsphere (Fig. \ref{fig.2}) but rather uses the same parameters as in \cite{microspherefabrication_OL}, not only for cross-validation but also to provide a clear evolution of the ray dynamics with the degree of deformation in microspheres.

The ray dynamics near the $y$-$z$ plane is regular in an ideal microsphere, see Fig. \ref{fig.4}(a), due to the conserved angular momenta $L_x$ and $L_z$ along the $x$- and $z$- axes, respectively. WGMs confined close to the surface of the sphere have $\sin\chi$ around 1.0. A slight quadrupolar deformation in the $x$-$y$ plane ($q=0.005$, stretched along the $y$-axis) hardly breaks the conservation of $L_z$ even over a very long time scale (\textcolor{red}{\textgreater}1,000,000 reflections); as a result, the ray dynamics remains regular around the $y$-$z$ plane, as shown in Fig. \ref{fig.4}(b). Figure \ref{fig.4}(c) shows the results when an additional quadrupolar deformation in the $y$-$z$ plane ($e=0.06$, stretched along the $z$ axis) is introduced which breaks the conservation of $L_x$. Nevertheless, the near-conservation of $L_z$ keeps the ray dynamics trivial near the $y$-$z$ plane, where chaotic diffusion emerges but remains well-confined between Kolmogorov-Arnold-Moser (KAM) tori \cite{1995PRL}. The KAM tori, in other words the smooth curves on the phase-space SOS, indicate confinement of the trajectory within the resonator. The ray dynamics represented by these points are quasi-periodical, implying stable behavior. Here, the ray behaves as if it were confined in a 2D chaotic resonator. Therefore, to observe true 3D chaotic dynamics, the conservation of $L_z$ must be significantly broken. To achieve this, a dipole deformation in the $x$-$y$ plane ($d=0.05$) is added, breaking the rotational symmetry around the $z$ axis and thus the conservation of $L_z$; the results are shown in Fig. \ref{fig.4}(d). Once the system becomes chaotic, scattered points with no clear pattern emerge on the Poincar\'e SOS. Initially, the chaotic diffusion is constrained by the KAM tori, much like in the case where $L_z$ is nearly conserved (i.e., Fig. \ref{fig.4}(c)). However, after a long time scale of about 600,000 reflections, although the KAM tori remain intact, the ray is no longer confined by them and instead freely and slowly accesses more of the real and phase spaces indicated by the randomly distributed points on the SOS (plotted for about 1,000,000 reflections). This slow process, which clearly depends on the nonconservation of $L_z$, is known as Arnold diffusion––a true 3D chaotic dynamic process \cite{arnol1964}. Our results show significant chaotic behavior in contrast to the results in \cite{microspherefabrication_OL} which we assume arises due to the level of deformation in all directions and the number of reflections considered (the scattered points become far more evident when we pass 600,000 reflections). It should be noted that the number of reflections in the ray trajectory calculations does not introduce additional chaotic behavior but instead provides a more complete representation of the trajectories that already exist within the chaotic regime. Increasing the number of reflections allows the system to explore a broader set of possible trajectories, thereby revealing the inherent richness of the chaotic dynamics without altering their fundamental nature. 

Figure \ref{fig.4}(e) shows the results when the deformation is further increased to match the experimental values measured in Fig. \ref{fig.2} and fitted in Appendix Fig. \ref{fig.App}. With this increased deformation, Arnold diffusion begins at around only 33,000 reflections. The onset of Arnold diffusion is strongly influenced by the degree of deformation: higher deformation modifies the phase-space structure, lowering the barriers for trajectories to escape from stable islands and enter chaotic regions. As a result, Arnold diffusion emerges at an earlier stage when the deformation is larger, leading to faster mixing across KAM tori. This earlier onset indicates that the system transitions to chaotic dynamics at lower deformation thresholds, enabling trajectories to escape from stable regions more readily. Such behavior enhances the sensitivity of the optical system to small perturbations and accelerates access to complex states. From an application perspective, this increased sensitivity may be exploited for high-resolution detection schemes, while the broader exploration of dynamic states could be harnessed for advanced light manipulation in photonic devices. 

At this point, we can see the advantage of having a full geometric quantification of a deformed 3D cavity via nondestructive means so as to ensure that the resonator can enter the chaotic regime, where interesting optical analogues of quantum tunnelling may appear, or highly-directional emission due to refractive escape may be observed. It is worth noting that establishing a universal quantitative criterion for identifying three-dimensional chaotic trajectories remains an open question in the field. Due to the current gap between theoretical progress and experimental validation, a generally accepted metric has yet to be defined. In this work, we contribute by replicating and extending previous results, revealing discrepancies and providing a broader set of scenarios, thereby offering new insights toward addressing this longstanding challenge.

\section{Conclusion}
\label{sect:sections}
While 2D chaotic microcavities have found increasing applications, the lack of structural characterization of 3D chaotic microcavities hinders the study of such complex physical systems. Here, X-ray $\mu$CT was demonstrated to fully access the geometrical information of 3D chaotic microcavities, enabling the accurate establishment of phase space chaotic ray dynamics in a deformed microsphere. X-ray $\mu$CT enables non-destructive imaging of 3D chaotic microcavities while maintaining submicron resolution. The advantage of non-destructive characterization is that it provides geometrical information about the sample before specific experimental testing for valuable calculations and simulation predictions, without relying on destructive measurements after the experiment. Due to the universality of the X-ray $\mu$CT technique, the proposed method can be easily extended to other complex photonic structures.

\appendix    

\section{Surface fitting results}
\label{sect:misc}
\begin{figure} [t]
\centering
      \includegraphics[width=0.7\textwidth]{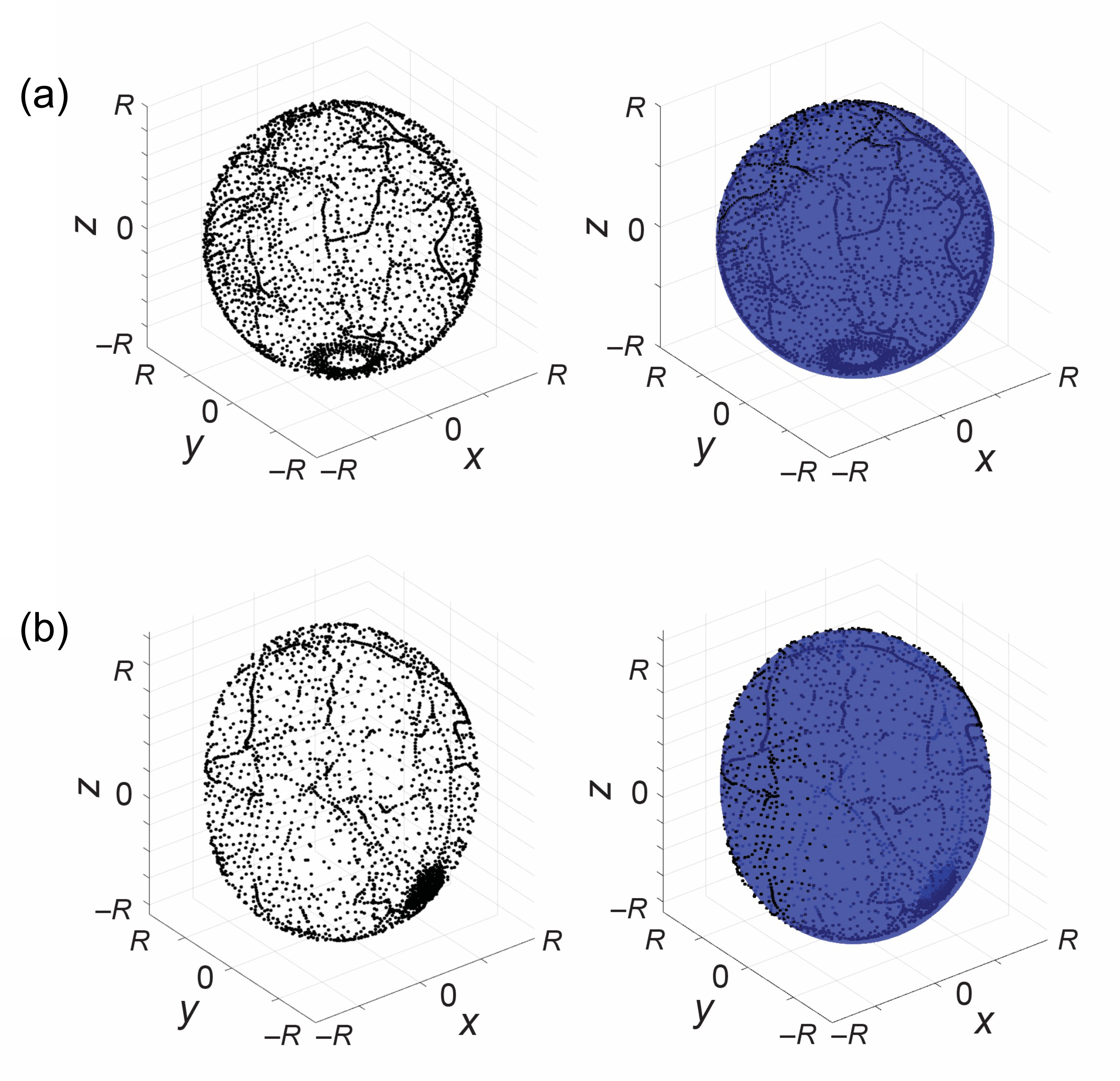}
  \caption{\textbf{Surface fitting results.} Extracted surfaces from X-ray $\mu$CT data in STL (stereolithography) format (left panel) and the fitting results using \eqref{Eq1} for a rotationally symmetric microsphere (a) and a deformed (asymmetric) microsphere (b) with a deformation of $\{q,e,d\}=\{0.02,0.08,0.05\}$.}
  \label{fig.App}
  
\end{figure}

\section*{Disclosures}
The authors declare no conflicts of interest.

\section* {Code, Data, and Materials Availability} 
Data underlying the results presented in this paper are not publicly available at this time but may be obtained from the authors upon reasonable request.

\section* {Acknowledgments}
The authors acknowledge support from the Okinawa Institute of Science and Technology Graduate University (OIST), the China Scholarship Council (CSC) (202306680004), and the Regional Innovation Strategy (RIS) through the National Research Foundation of Korea (NRF) funded by the Ministry of Education (MOE) (2021RIS-001, 2022R1C1C1006040). The authors would like to thank the Engineering Section, the Scientific Computing \& Data Analysis Section, and the Scientific Imaging Section of OIST for technical assistance. SNC acknowledges useful discussions with M. Hentschel. S. Li, K. Tian, and M. Z. Jalaludeen acknowledge support from the Japan Society for the Promotion of Science (JSPS) KAKENHI through Grant-in-Aid for Scientific Research (C) (23K04617), Grant-in-Aid for Early-Career Scientists (22K14621), and Grant-in-Aid for JSPS fellows (25KJ2244).

\section*{Supplementary material}
See the supplementary material for the X-ray characterization results of the 3D chaotic microcavity.


\bibliography{References}   
\bibliographystyle{spiejour}   

\vspace{1ex}

\end{spacing}
\end{document}